\documentclass[epj]{svjour}

\usepackage{graphics}
\usepackage{dcolumn}
\usepackage{bm}
\usepackage{amsmath} 
\usepackage{hyperref}
\usepackage{amsfonts}
\usepackage{wasysym}
\usepackage{epsfig}    
\begin{document}
\title{Domain wall delocalization, dynamics and fluctuations in an exclusion process with two internal states}
\titlerunning{Domain wall delocalization in an exclusion process with two internal states}
\author{ Tobias Reichenbach \and Thomas Franosch \and Erwin Frey
}                     
%
%
\institute{
Arnold Sommerfeld Center for Theoretical Physics (ASC) and
  Center for NanoScience (CeNS), Department of Physics,
  Ludwig-Maximilians-Universit\"at M\"unchen, Theresienstrasse 37,
  D-80333 M\"unchen, Germany} 

\date{Received: date / Revised version: date}
%
\abstract{We investigate the delocalization transition appearing in an exclusion process with two internal states resp. on two parallel lanes. At the transition, delocalized domain walls form in the density profiles of both internal states, in agreement with a mean-field approach. Remarkably, the topology of the system's phase diagram allows for the delocalization of a (localized) domain wall when approaching the transition. We quantify the domain wall's delocalization close to the transition by analytic results obtained within the framework of the domain wall picture. Power-law dependences of the domain wall width on the distance to the delocalization transition as well as on the system size are uncovered, they agree with numerical results.
} 
\maketitle

\section{Introduction}
   
Driven one-dimensional transport phenomena~\cite{SchmittmannZia} currently receive much attention as they constitute a challenging class of non-equilibrium dynamical systems. In such systems, collective effects induce unexpected phenomena,
including e.g. boundary-induced phase transitions~\cite{krug-1991-76} or  pattern formation~\cite{georgiev-2005-94}. 
Possible applications are  found in a large variety of contexts, ranging from biology (e.g. the motion of ribosomes along mRNA~\cite{macdonald-1968-6} or molecular motors on intracellular filaments~\cite{hirokawa-1998-279,Howard,lipowsky-2001-87,kruse-2002-66,klumpp-2003-113,klein-2005-94,Hinsch}) to electron hopping transport with applied voltage~\cite{zutic-2004-76,reichenbach-2006-97} and vehicular highway traffic~\cite{helbing-2001-73,chowdhury-2000-329}.

As unifying descriptions of such non-equilibrium systems are still lacking, much effort is devoted to the understanding of particular models and the identification of universal properties as well as analytic methods. In this context, the Totally Asymmetric  Exclusion Process (TASEP) has emerged as a paradigm (see e.g.~\cite{derrida-1998-301,Schutz} for a review). There, particles move unidirectionally from left to right through a one-dimensional lattice. The injection resp. extraction rates at the left resp. right boundaries serve as control parameters; tuning them, different phases of the stationary-state density are observed. Although being exactly solvable~\cite{derrida-1992-69,schuetz-1993-72,derrida-1993-26}, a mean-field (MF) approach already yields the exact phase diagram, originating in an exact MF current-density relation~\cite{derrida-1992-69}. Beyond MF, fluctuations have been successfully taken into account in the framework of the domain wall picture~\cite{kolomeisky-1998-31,santen-2002-106}. There, one considers a domain wall separating a low-density (LD) region from a high-density (HD) one. The domain wall's  dynamics yields information about the phase behavior and boundary effects arising in finite systems.

\begin{figure}[htbp]
\begin{center}
\includegraphics[scale=1]{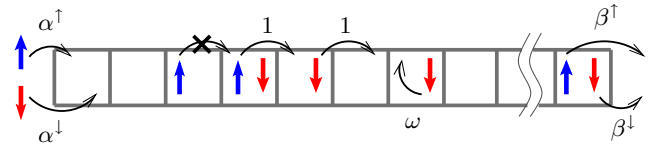}
\caption{(Color online) Illustration of an exclusion model with
two internal states, denoted as spin-up ($\uparrow$) and spin-down ($\downarrow$). Particles are injected at the left boundary at rates $\alpha^\uparrow, \alpha^\downarrow$, and extracted at the right at rates $\beta^\uparrow,\beta^\downarrow$.
Within bulk, particles move unidirectionally to the right, or flip their state (at rate $\omega$), always obeying Pauli's exclusion principle.
\label{cartoon_spin}}
\end{center}
\end{figure}
Recently, we have proposed a generalization of TASEP where particles possess two weakly coupled internal states \cite{reichenbach-2006-97,reichenbach-2007-9}. As an example, these internal states may correspond to different parallel lanes in vehicular traffic on highways~\cite{helbing-2001-73}. Concerning intracellular transport, microtubules consists of typically 12-14 parallel lanes, and molecular motors progressing on them may (though rarely) switch between these lanes. Our work provides a minimal model that takes such lane changes into account. Also, the internal states may describe spin states of electrons, e.g., when performing hopping transport through a chain of quantum dots~\cite{zutic-2004-76}, suggesting possible application of our model to spintronics devices.

In the subsequent paper, we  adopt the language of spins for the internal states. The system's dynamics, described in the following, is depicted in Fig.~\ref{cartoon_spin}. Particles with spin-up (down) state enter at the left boundary at rate $\alpha^\uparrow$ ($\alpha^\downarrow$), under the constraint of Pauli's exclusion principle. The latter means that each lattice site may at most be occupied by one particle of a given state, such that particles with spin-up and spin-down may share one site, but no two spin-up or spin-down particles are allowed. (In the context of multi-lane traffic, this translates into simple site exclusion.) Within the lattice, particles hop to the neighboring right lattice site at constant rate (which we set to unity), again respecting Pauli's exclusion principle. Spin flip events may occur, we denote the respective rate by $\omega$. Finally, particles are extracted at the right boundary at rate $\beta^\uparrow$ ($\beta^\downarrow$), depending on their spin state.  
   
\begin{figure}  
\begin{center}                     
\includegraphics[scale=1]{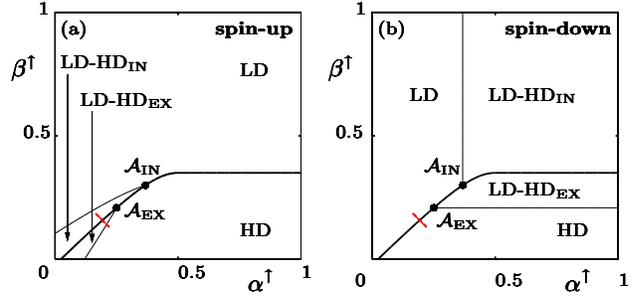}
\caption{Phase diagrams for the density of spin-up (a) and spin-down (b): two-dimensional cuts for fixed values of $\alpha^\downarrow=0.25$, $\beta^\downarrow=0.3$ and in the asymptotic limit of large $\Omega$. Hereby, $\Omega=\omega/L$ is the gross spin flip rate.
Lines of continuous transitions (thin) intersect the delocalization transition line (bold) in multicritical points $\mathcal{A}_\text{IN}$ and  $\mathcal{A}_\text{EX}$. The scope of this paper is to elucidate the system's behavior at the delocalization transition as well as when approaching the latter along the path shown in red (or grey).  \label{phasediag_ex}} 
\end{center}
\end{figure}

In~\cite{reichenbach-2006-97,reichenbach-2007-9} we have taken advantage of a MF approach in a continuum limit, see e.g.~\cite{Mukamel} for a review, to describe the emerging non-equilibrium stationary density profiles. Comparison to data from stochastic simulations has, through finite size scaling, uncovered an apparent exactness of the analytic results in the limit of large system sizes. We have traced back this exactness to the weak coupling of the two internal states, see below, as well as the exactness of the MF approach for TASEP~\cite{derrida-1998-301}.  As it implies the exactness of the analytically derived phase diagrams (see Fig.~\ref{phasediag_ex} for a two-dimensional cut), we have full knowledge of the system's phase behavior. The latter exhibits a rich variety of phases, in particular, a localized domain wall, separating a low-density (LD) from a high-density (HD) region, may emerge in the density profile of one spin state. We refer to this situation as a coexistence of LD and HD, abbreviated as LD-HD. Continuous as well as discontinuous transitions between the different phases appear and induce multicritical points (e.g. $\mathcal{A}_\text{IN}$ and $\mathcal{A}_\text{EX}$ in Fig.~\ref{phasediag_ex}).
Hereby, as we already announced in~\cite{reichenbach-2006-97,reichenbach-2007-9}, domain wall delocalization appears at the discontinuous transition. Namely, as an example, consider the red (grey) line in Fig.~\ref{phasediag_ex}. It crosses a discontinuous transition, which seperates two LD-HD phases. In both, localized domain walls emerge, but at different positions in bulk. Upon crossing the delocalization transition along the red line, the domain wall delocalizes, and then localizes again at a different position. The aim of this article is the investigation of the domain wall's delocalization process when approaching the discontinuous transition.

In this paper, we present detailed investigations on the delocalization transition. Starting from MF considerations, we  show that at the transition line delocalized domain walls appear in the density profiles of both spin states. More precise, the MF approach predicts a one-parameter family of solutions with domain walls for spin-up and spin-down, which we conclude to perform  coupled random walks. Further on, including fluctuations via the domain wall picture  allows to quantify  the delocalization of a localized domain wall when approaching the delocalization transition, e.g. along the path depicted in red (or grey) in Fig.~\ref{phasediag_ex}.
   
The outline of this paper is the following. In the next section we review the MF approach (see our previous article~\cite{reichenbach-2007-9} for a detailed discussion), and present the solution for the density profiles at the delocalization transition line. The solution turns out to be not unique, but is a one-parameter family, indicating the delocalization. Our stochastic simulations are described in Section~\ref{stoch_sim}. Section~\ref{domain_wall_pict} goes beyond MF by taking fluctuations into account. Within the domain wall picture, we investigate the delocalization of a domain wall when approaching the delocalization transition. As a result, we find power law dependences for the width of the domain wall distribution on the system size and the distance to the delocalization transition line. Our conclusions are presented in
Sec.~\ref{concl}.

\section{Predictions from the mean-field approach}
\label{phase_coex1}

We are interested in the stationary density profiles which emerge in the system's non-equilibrium steady state. 
Therefore, denote  the occupation number of site $i$ for spin-up resp. spin-down state by $n_i^{\uparrow}$ resp. $n_i^{\downarrow}$, i.e. $n_i^{\uparrow(\downarrow)}\in\{0,1\}$, depending on whether this site is occupied by a particle with corresponding spin state or not.
Performing sample averages, we obtain the average occupation, $\rho_i^{\uparrow(\downarrow)}\equiv\langle n_i^{\uparrow(\downarrow)}\rangle$.  In the mean-field (MF) approximation, higher order
 correlations
between the occupation numbers are neglected, i.e. we assume
\begin{equation}
\langle n_i^r n_j^s\rangle=\rho_i^r\rho_j^s \,; \quad r,s\in\{\uparrow,\downarrow\} \, .
\end{equation}
The dynamical rules of the system lead to equations for the densities in the stationary state; see Ref.~\cite{reichenbach-2007-9} for a detailed discussion. Further on, in a continuum limit,  the total length of the lattice is set to unity and the limit of number of lattice sites $L\rightarrow\infty$ is considered. Hereby, the densities $\rho_i^{\uparrow (\downarrow)}$ approximate smooth functions $\rho^{\uparrow(\downarrow)}(x)$ with $x\in[0,1]$; for the latter, we eventually obtain two coupled differential equations:
\begin{equation}
\partial_x j^\uparrow=\Omega[\rho^\downarrow-\rho^\uparrow]\,
,\quad \partial_x j^\downarrow=\Omega[\rho^\uparrow-\rho^\downarrow]\,.
\label{cont_curr}  
\end{equation}
Here,  we have defined currents for the individual spin states: $j^{\uparrow(\downarrow)}(x)=\rho^{\uparrow(\downarrow)}(x)\big[1-\rho^{\uparrow(\downarrow)}(x)\big]$. Also, the gross spin flip rate $\Omega=\omega L$ was introduced; in a mesoscopic scaling, we keep this rate fixed when performing the limit $L\rightarrow\infty$, see Refs.~\cite{parmeggiani-2003-90,parmeggiani-2004-70}. In this way, competition between the bulk processes (spin flips) and the boundary processes (particle injection and extraction) emerges. This weak coupling is opposed to strong coupling, where $\omega$ is kept constant when considering large systems, which leads to different effects~\cite{schmittmann-2005-70,pronina-2004-37,pronina-2006-372}. A closed analytic form for the solution to Eqs.~(\ref{cont_curr}) is feasible and has been presented in~\cite{reichenbach-2007-9}.

The injection and extraction processes lead to boundary conditions for Eqs.~(\ref{cont_curr}):  
\begin{align}
\rho^\uparrow(0)=&~\alpha_{\text{eff}}^\uparrow \,,\cr
\rho^\downarrow(0)=&~\alpha_{\text{eff}}^\downarrow \,,\cr
\rho^\uparrow(1)=&~1-\beta_{\text{eff}}^\uparrow \,,\cr
\rho^\downarrow(1)=&~1-\beta_{\text{eff}}^\downarrow   \, , \label{bound_cond}
\end{align}
where we have introduced effective rates $\alpha_{\text{eff}}^{\uparrow(\downarrow}), \beta_{\text{eff}}^{\uparrow(\downarrow)}$ according to 
\begin{align}
\alpha_{\text{eff}}^{\uparrow(\downarrow)}&=\text{min}\big[\alpha^{\uparrow(\downarrow)},\frac{1}{2}\big]\,,  \cr
\beta_{\text{eff}}^{\uparrow(\downarrow)}&=\text{min}\big[\beta^{\uparrow(\downarrow)},\frac{1}{2}\big]  \, .
\end{align}
They reflect the fact that bulk processes limit the individual spin currents to maximal values of $1/4$, corresponding to densities of $1/2$. Injection or extraction rates exceeding this value cannot lead to larger currents, but effectively act as $1/2$ (see~\cite{reichenbach-2007-9}).

The boundary conditions~(\ref{bound_cond}) apparently overdetermine the system of two differential equations~(\ref{cont_curr}). Indeed, two types of singularities may occur where the densities exhibit discontinuities: boundary layers (discontinuity at the boundary) and domain walls (discontinuity in bulk).  An analytic description of the domain wall positions is feasible from the observation that the spin currents $j^{\uparrow(\downarrow)}(x)=\rho^{\uparrow(\downarrow)}(x)\big[1-\rho^{\uparrow(\downarrow)}(x)\big]$ at these positions must be continuous. The only solution to this condition is that a density $\rho^{\uparrow(\downarrow)}$ changes to the new value $1-\rho^{\uparrow(\downarrow)}$ at the position of the discontinuity. 

As particles cannot leave the system in bulk, the particle current $J=j^\uparrow+j^\downarrow$ is spatially conserved. In case that no boundary layer occurs at the left, i.e. the densities do not have any discontinuities there, it is given from~(\ref{bound_cond}) by the value $J_\text{IN}=\alpha_\text{eff}^\uparrow(1-\alpha_\text{eff}^\uparrow)+\alpha_\text{eff}^\downarrow(1-\alpha_\text{eff}^\downarrow)$. In this case, the system is dominated by injection processes; we refer to the region in parameter space, where this behavior occurs, as the IN-region. On the other hand, also extraction can determine the system's behavior, the particle current is then given by $J_\text{EX}=\beta_\text{eff}^\uparrow(1-\beta_\text{eff}^\uparrow)+\beta_\text{eff}^\downarrow(1-\beta_\text{eff}^\downarrow)$, occurring in the EX-region.

The \emph{delocalization transition} occurs at $J_\text{IN}=J_\text{EX}$. There, the system is in a superposition of the injection dominated and the extraction dominated behavior. In this situation, delocalized domain walls appear in the density profiles of \emph{both} spin states. Indeed, only when $J_\text{IN}=J_\text{EX}$, density profiles satisfying the boundary conditions at the left boundary as well as at the right one are feasible. Instead of discontinuities at the boundaries, they exhibit discontinuities within bulk: domain walls.
In the following, we want to present the MF picture for the density behavior at the delocalization transition and show that a one-parameter family of solutions emerges within the MF approach.

\subsection{The one-parameter family of analytic solutions}

\begin{figure}
\begin{center}
\includegraphics[scale=1]{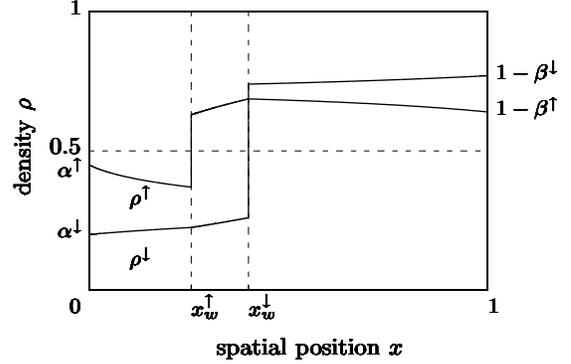}
\caption{Domain walls in the densities of both spin states at the delocalization transition: results from the MF approach.  The positions of the domain walls are only determined up to one degree of freedom: choosing the position $x_w^\uparrow$ for the domain wall of spin-up  yields the  position $x_w^\downarrow$ for spin-down and vice versa. The densities found in stochastic simulations are  the average over this one-parameter family of solutions. Parameters are $\alpha^\uparrow=0.45, \alpha^\downarrow=0.2, \beta^\uparrow=0.36, \beta^\downarrow=0.23$ and $\Omega=0.3$.  \label{coex_dens}}
\end{center}
\end{figure}
The generic picture of the MF analytic solution at the delocalization transition is presented in Fig.~\ref{coex_dens}. As the equation $J_\text{IN}=J_\text{EX}$ is fulfilled, the density profiles can match the left as well as the right boundary conditions.  This implies that, in bulk, domain walls arise at positions $x_w^\uparrow$ and $x_w^\downarrow$ in spin-up resp. spin-down state;  we have chosen $x_w^\uparrow<x_w^\downarrow$, the other case is obtained by the symmetry of the two spin states.
In the vicinity of the left boundary,
the densities are given by the analytic solution $\rho^{\uparrow,\downarrow}_l$ obeying the left boundary conditions $\rho^\uparrow_l(x=0)=\alpha^\uparrow_{\text{eff}}$, $\rho^\downarrow_l(x=0)=\alpha^\downarrow_{\text{eff}}$. At the point $x_w^\uparrow$, the density of spin-up jumps from the value $\rho^\uparrow_l(x_w^\uparrow)$ to the value $1-\rho^\uparrow_l(x_w^\uparrow)$, while the density of spin-down is continuous. To the right of $x_w^\uparrow$, the densities follow a different branch of analytic solution, we name it $\rho^{\uparrow,\downarrow}_m$ as the intermediate branch. It is determined by the boundary conditions $\rho^\uparrow_m(x_w^\uparrow)=1-\rho^\uparrow_l(x_w^\uparrow)$ and $\rho^\downarrow_m(x_w^\downarrow)=\rho^\downarrow_l(x_w^\uparrow)$. At the point $x_w^\downarrow$, the density of spin-down changes discontinuously from a value $\rho^\downarrow_m(x_w^\downarrow)$ to a value $1-\rho^\downarrow_m(x_w^\downarrow)$, while the density of spin-up remains continuous. To the right of $x_w^\downarrow$, the densities obey the analytic solution $\rho^{\uparrow,\downarrow}_r$ satisfying the boundary conditions at the right: $\rho^\uparrow_r(x=1)=1-\beta^\uparrow_{\text{eff}}$, $\rho^\downarrow_r(x=1)=1-\beta^\downarrow_{\text{eff}}$.\\
Now, for the situation to be feasible, we have  additional conditions coming from considering the continuity of the spin currents at the point $x_w^\downarrow$, namely $\rho_r^\uparrow(x_w^\downarrow)=\rho^\uparrow_m(x_w^\downarrow)$ and $\rho^\downarrow_r(x_w^\downarrow)=1-\rho^\downarrow_m(x_w^\downarrow)$. As the particle current is conserved, these two conditions are not independent: the validity of one implies the validity of the other.  Only one condition thus exists for two variables, the latter being the positions of the domain walls $x_w^\uparrow$ and $x_w^\downarrow$. Our considerations therefore leave us with a one-parameter family of solutions; as parameter, we can e.g. take one of the domain wall positions, $x_w^\uparrow$ or $x_w^\downarrow$.

\subsection{Spatial boundaries for the domain walls}

\begin{figure}
\begin{center}
\includegraphics[scale=1]{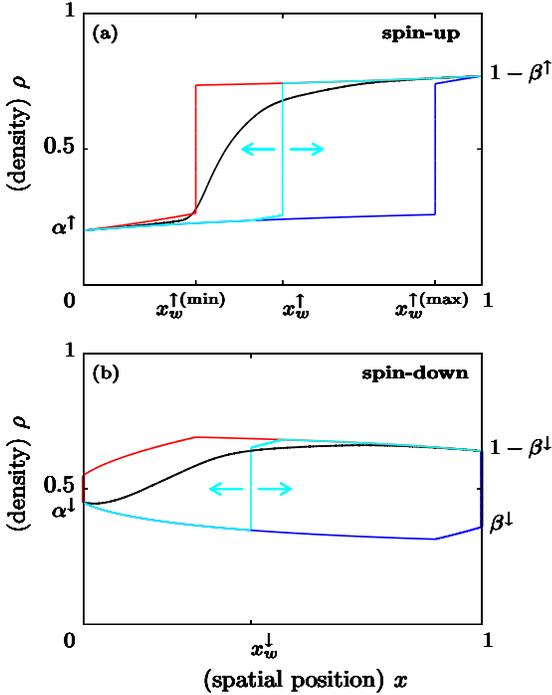}
\caption{(Color online) Analytic versus numerical results for the density profiles at the delocalization transition. The analytic approach predicts a one-parameter family of solutions, we show the one with the minimal domain wall positions (red or grey) and with the maximal one (blue or dark grey) as well as an intermediate (light blue or light grey). Stochastic simulations yield an average over these solutions, resulting in densities interpolating between the extrema (black). Parameters are the same as in Fig.~\ref{coex_dens}.
\label{coex_rev}}
\end{center}
\end{figure}

In the course of time, the system takes all states specified by the one-parameter family of solutions. Averaged over sufficiently long times, the densities represent the average over the latter, resulting in  a smoothening of the
density profiles. In Fig.~\ref{coex_rev}, we show how the results may look like. We observe that certain spatial boundaries ($x_w^{\text{(min)}}$ and $x_w^{\text{(max)}}$ in the above picture) exist. They define  the bulk region where the domain walls may appear. The description of these boundaries is the scope of this subsection.

The boundaries for the residence regions of the domain walls originate in the constraint
that the domain wall positions must lie between $0$ and $1$. Indeed, let us   consider the minimal positions $x_w^{\uparrow\text{(min)}}, x_w^{\downarrow\text{(min)}}$ which are feasible for  the domain walls in the spin-up and spin-down state. We had found in Subsec.~\ref{phase_coex1} that fixing the domain wall position in the density profile of one of the states determines the position of the other domain wall. The situation $x_w^{\downarrow\text{(min)}}= x_w^{\uparrow\text{(min)}}=0$ can thus only emerge if $x_w^{\downarrow\text{(min)}}=0$ induces $x_w^{\uparrow\text{(min)}}=0$ (corresponding to the multicritical point $\mathcal{A}_\text{IN}$, see the detailed discussion in Ref.~\cite{reichenbach-2007-9}). In general, only one of these minimal domain wall positions is at $0$, and induces a minimal position for the other one which is larger than $0$. Analogously, in general, a maximal domain wall position of $1$ only occurs for one of the spin states, and induces a domain wall position smaller than $1$ for the other. The exceptional case $x_w^{\downarrow\text{(max)}}= x_w^{\uparrow\text{(max)}}=1$ occurs at the multicritical point $\mathcal{A}_\text{EX}$, as the latter lies at the transition from a coexistence phase to a HD phase, for both spin states, see Ref.~\cite{reichenbach-2007-9}

Which situation occurs, may be read off from the phase diagrams by considering the phases in the vicinity of the delocalization transition. As an example, we consider a point on the delocalization transition line in Fig.~\ref{phasediag_ex} (a), (b), at the intersection with the path shown in red. In its vicinity, in the EX-region, the density of spin-down is in a pure HD-phase, such that a domain wall at position $x_w^\downarrow=0$ forms there when approaching the delocalization transition. The density of spin-up exhibits a localized shock at a position $0<x_w^\uparrow<1$. We thus observe $x_w^{\downarrow\text{(min)}}=0$ and
$0<x_w^{\uparrow\text{(min)}}<1$. Similarly, considering the IN-region in the vicinity of this point yields the maximal feasible domain wall positions there:  $x_w^{\downarrow\text{(max)}}=1$ and
$0<x_w^{\uparrow\text{(max)}}<1$.

The analytic solutions to the density profiles in  the situation of $x_w^{\uparrow\text{(min)}}, x_w^{\downarrow\text{(min)}}$ (red) as well as for $x_w^{\uparrow\text{(max)}}, x_w^{\downarrow\text{(max)}}$ (blue) are shown in Fig.~\ref{coex_rev}. The intermediate solution (light blue) varies between these two extrema.
In the system, averaged over sufficiently long time, this results in smoothened density profiles. In Fig.~\ref{coex_rev} we show the results from stochastic simulations as black lines, agreeing with our analytic considerations.

\section{Stochastic simulation methods}

\label{stoch_sim}   

To validate our analytic calculations, we have carried out extensive stochastic simulations. An efficient simulation method originally due to Gillespie~\cite{gillespie-1976-22,gillespie-1977-81} was implemented. There, in each time step, a random number determines whether particle injection, particle extraction, a spin flip event or a particle hopping forward may occur in the next time step. The time interval to the next process is chosen from an exponential waiting time distribution.

Simulations in the neighborhood of the delocalization transition need long waiting times  until the densities reach stationary profiles, due to the random walks performed by the domain walls. We carried out simulations with up to $10^8$ time windows, each consisting of $10\times L$ steps of updating. An example of resulting density profiles is given in Fig.~\ref{coex_rev} for a point on the delocalization transition line,  the densities are observed to interpolate between two extremal analytic solutions.

\begin{figure}
\begin{center}
\includegraphics[scale=1]{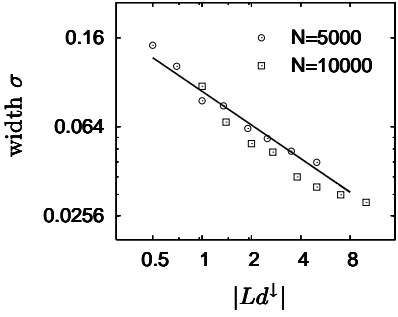}
\caption{The width $\sigma$ of the domain wall distribution in the spin-up density profile for the case of a localized domain wall there, at a small distance $d^\uparrow$ to the delocalization transition. Numerical results are shown in a double logarithmic scale, depending on $|Ld^\downarrow|$. The data are obtained from system sizes of $L=5000$ (open circles) and $L=10000$ (open boxes), with $d^\downarrow$ varied between $1\times10^{-4}$ and $1\times 10^{-3}$. The solid line indicates the slope $-\frac{1}{2}$. Approximately, we recover $\sigma\sim|Ld^\downarrow|^{-1/2}$.
 \label{scaling}}
\end{center}
\end{figure}

In the subsequent section, we investigate the scaling of the width $\sigma$ of the domain wall when approaching the delocalization transition. For the numerical data, we have computed the stationary density profiles for system sizes of $L=5000$ and $L=1000$ at different distances to the delocalization transition. The width $\sigma$ was obtained by fitting the densities profiles in the vicinity of the domain wall with the function $A\cdot\text{erf}[(x-b)/\sigma]+C+Dx$, with parameters $A,b,C,D,\sigma$. The first term describes a Gaussian distribution of the domain wall, while the latter two terms account for the background density profile (to linear order in $x$).
The results are shown in Fig.~\ref{scaling} and agree well with the analytic predictions. We attribute deviations to the fact that for large $\sigma$ the above fitting becomes less accurate as the domain wall distribution deviates from a gaussian.

\section{Beyond mean-field: Approaching the delocalization transition}

\label{domain_wall_pict}

In the previous Sections, we have described how delocalized domain walls in the density profiles of both states  appear at the delocalization transition. Here, we investigate the emergence of delocalization when approaching the delocalization line
by applying the domain wall picture. Including fluctuations, the latter allows us to calculate the width of the domain wall distribution
for finite system sizes $L$. It diverges as one approaches the delocalization line, yielding delocalized domain walls.

\subsection{The domain wall picture}
 
The domain wall picture~\cite{kolomeisky-1998-31} allows to include fluctuations, such as the particle-number fluctuations in the TASEP~\cite{santen-2002-106}, helping in understanding its dynamics~\cite{pierobon-2005-72}, and thus go beyond the MF approximation.
In this approach, we start with the assumption that sharp and localized domain walls appear in the density profiles of both states. Injection and extraction of  particles as well as spin-flips induce dynamics for them, they  perform random walks.
Our focus is on the emerging stationary state. As the latter arises in a non-equilibrium system, it need not obey detailed balance.
Writing down the master equation for the random walk, we are able to calculate the fluctuations of the domain wall positions around the stationary values, yielding the width of the domain walls or boundary layers which form in finite systems.

The dynamics induced by entering and exiting particles or spin-flips is the following. If the total number of particles of given internal state in the system is increased by one,  the domain wall in the density profile of this state moves a certain value to the left. In the other case, if the  number of particles of given state is decreased by one,  the domain wall of this state moves to the right.

Denote $\Delta_\uparrow$ resp. $\Delta_\downarrow$ the height of the domain wall in the density profile of the spin-up resp. spin-down states. They are functions of the domain wall positions: $\Delta_\uparrow=\Delta_\uparrow(x_w^\uparrow,x_w^\downarrow),~\Delta_\downarrow=\Delta_\downarrow(x_w^\uparrow,x_w^\downarrow)$.  Increasing the number of spin-up particles by one shifts the position $x_w^\uparrow$  by the value $\delta x_w^\uparrow=-L^{-1}\Delta_\uparrow^{-1}$, and decreasing its number   results in a shift $\delta x_w^\uparrow=L^{-1}\Delta_\uparrow^{-1}$. The changes $\delta x_w^\downarrow$ result analogously from particles with spin-down entering and exiting.

\subsection{Fokker-Planck equation and the generic form of the domain wall distribution \label{subsec_fokker_planck}}

In our model, six processes alter the particle number of one or both internal states:
\begin{verse}
(i) A particle with spin-up entering  at the left boundary\\
(ii) A particle with spin-up leaving  at the right boundary\\
(iii) A particle with spin-down entering   at the left boundary\\
(iv) A particle with spin-down leaving at the right boundary\\
(v) A particle with spin-up flipping to spin-down in  bulk\\
(vi) A particle with spin-down flipping to spin-up in  bulk\\
\end{verse}

Their probabilities and resulting changes $\delta x_w^\uparrow,~\delta x_w^\downarrow$ are listed in Tab.~\ref{random_walk}.

 \begin{table}
\begin{tabular}{|c|c|c|c|} \hline
Process & Probability $\mathcal{W}$ & $\delta x_w^\uparrow$  & $\delta x_w^\downarrow$  \\ \hline \hline
(i) & $\alpha_\text{eff}^\uparrow(1-\alpha_\text{eff}^\uparrow)$ & $-L^{-1}\Delta_\uparrow^{-1}$&0  \\ \hline
(ii) & $\beta_\text{eff}^\uparrow(1-\beta_\text{eff}^\uparrow)$ & $L^{-1}\Delta_\uparrow^{-1}$&0  \\ \hline
(iii) & $\alpha_\text{eff}^\downarrow(1-\alpha_\text{eff}^\downarrow)$ & 0&$-L^{-1}\Delta_\downarrow^{-1}$  \\ \hline
(iv) & $\beta_\text{eff}^\downarrow(1-\beta_\text{eff}^\downarrow)$ &0& $L^{-1}\Delta_\downarrow^{-1}$  \\ \hline
(v) & $\Omega\int_0^1\rho^\uparrow(x)[1-\rho^\downarrow(x)]dx$ & $L^{-1}\Delta_\uparrow^{-1}$&$-L^{-1}\Delta_\downarrow^{-1}$  \\ \hline
(vi) & $\Omega\int_0^1\rho^\downarrow(x)[1-\rho^\uparrow(x)]dx$ & $-L^{-1}\Delta_\uparrow^{-1}$&$L^{-1}\Delta_\downarrow^{-1}$  \\ \hline
\end{tabular}
\caption{Six processes lead to changes in the domain wall positions.  This table lists their probabilities and resulting changes $\delta x_w^\uparrow,~ \delta x_w^\downarrow$.
\label{random_walk}}
\end{table}

To shorten our expressions, we introduce the vectors $\underline{x}_w=(x_w^\uparrow,x_w^\downarrow)$ and $ \delta\underline{ x}_w=(\delta x_w^\uparrow,\delta x_w^\downarrow)$. Defining $P(\underline{x}_w,t)$ as the probability density of finding the domain wall positions $x_w^\uparrow,~x_w^\downarrow$ at time t, the processes (i)-(vi) allow us to write down the master equation
\begin{align}
\partial_t P(\underline{x}_w,t)=\sum_{\delta\underline{x}_w}\Big\{& P(\underline{x}_w+\delta\underline{x}_w,t)\mathcal{W}(\underline{x}_w+\delta\underline{x}_w\rightarrow \underline{x}_w)   \cr
& -P(\underline{x}_w,t)\mathcal{W}(\underline{x}_w\rightarrow \underline{x}_w+\delta\underline{x}_w)\Big\}
 \, .
\label{master}
\end{align}

The transition probabilities $\mathcal{W}(\underline{x}_w\rightarrow \underline{x}_x+\delta \underline{x}_w)$
are the ones listed in Tab.~\ref{random_walk}.

For the expectation values of the changes $\delta x_w^\uparrow,~\delta x_w^\downarrow$ we obtain
\begin{align}
\langle\delta x_w^\uparrow\rangle=&L^{-1}\Delta_\uparrow^{-1}\big\{\beta_\text{eff}^\uparrow(1-\beta_\text{eff}^\uparrow)-\alpha_\text{eff}^\uparrow(1-\alpha_\text{eff}^\uparrow)\cr
&\quad+\Omega\int_0^1[\rho^\uparrow(x)-\rho^\downarrow(x)]dx\big\}  \, ,  \cr
\langle\delta x_w^\downarrow\rangle=&L^{-1}\Delta_\downarrow^{-1}\big\{\beta_\text{eff}^\downarrow(1-\beta_\text{eff}^\downarrow)-\alpha_\text{eff}^\downarrow(1-\alpha_\text{eff}^\downarrow)\cr
&\quad+\Omega\int_0^1[\rho^\downarrow(x)-\rho^\uparrow(x)]dx\big\}    \, ,
\end{align}
which, of course, depend on the densities $\rho^\uparrow$ and $\rho^\downarrow$ and thereby on the domain wall positions $x_w^\uparrow,~x_w^\downarrow$. Investigating the fixed point of the random walk, i.e. the values $\bar{x}_w^\uparrow,~\bar{x}_w^\downarrow$ where  $\langle\delta x_w^\uparrow\rangle=\langle\delta x_w^\downarrow\rangle=0$, we find as a necessary condition $\alpha_\text{eff}^\uparrow(1-\alpha_\text{eff}^\uparrow)+\alpha_\text{eff}^\downarrow(1-\alpha_\text{eff}^\downarrow)=\beta_\text{eff}^\uparrow(1-\beta_\text{eff}^\uparrow)+\beta_\text{eff}^\downarrow(1-\beta_\text{eff}^\downarrow)$, or $J_\text{IN}=J_\text{EX}$, describing the delocalization transition. Thus, domain walls within the density profiles of \emph{both} states are only feasible there; otherwise, at most one of the domain wall positions, say $\bar{x}_w^\uparrow$, can lie inside the bulk. The other one, say $\bar{x}_w^\downarrow$, is driven outside of the system and turns into a boundary layer. We then set $\bar{x}_w^\downarrow=0$ or $\bar{x}_w^\downarrow=1$, depending on whether a HD or a LD phase occurs. With this convention, the values $\bar{x}_w^\uparrow,~\bar{x}_w^\downarrow$ are given by the MF analytic solution.

We are interested in the fluctuations of the domain wall positions around the mean values $\bar{x}_w^\uparrow,~\bar{x}_w^\downarrow$, and therefore define the quantities
\begin{align}
y^\uparrow=&x_w^\uparrow-\bar{x}_w^\uparrow\, ,\cr
y^\downarrow=&x_w^\downarrow-\bar{x}_w^\downarrow \, ,\cr
\end{align}
as the deviations. Again, they are arranged in a vector $\underline{y}=(y^\uparrow,y^\downarrow)$, and we have $\delta\underline{y}=\delta\underline{x}_w$.

Applying the Kramers-Moyal expansion~\cite{taeuber-notes} of the master equation~(\ref{master}) around the mean-field values to second order in the quantities $y^\uparrow,~y^\downarrow$ results in the Fokker-Planck equation
\begin{equation}
\partial_tP(\underline{y},t)=-\partial_i[a_i(\underline{y})P(\underline{y},t)]+\frac{1}{2}\partial_i\partial_j[B_{ij}(\underline{y})P(\underline{y},t)]\,.
\label{fokker_planck}
\end{equation}
Here, the indices $i,j$ stand for spin-up and spin-down.  In the above equation, the summation convention implies summation over them. The partial derivative $\partial_i$ is the short-hand notation of $\partial/\partial y^i$.

The quantities $a_i$ and $B_{ij}$ are, according to the Kramers-Moyal expansion:
\begin{align}
a_i(\underline{y})=&\sum_{\delta\underline{y}} \delta y^i\mathcal{W}(\underline{y}\rightarrow \underline{y}+\delta \underline{y})  \,, \cr
B_{ij}(\underline{y})=&\sum_{\delta \underline{y}}\delta y^i \delta y^j\mathcal{W}(\underline{y}\rightarrow \underline{y}+\delta\underline{y}) \,.
\label{a_B}
\end{align}
For our case, they are given in App.~\ref{app_fokker_planck}.

Now, we expand the Fokker-Planck equation around $\underline{y}=0$, using the linear noise approximation of van Kampens $\Omega$-expansion~\cite{VanKampen}.  As $B_{ij}(\underline{y}=0)\neq 0$, we consider approximately $B_{ij}(\underline{y})\approx B_{ij}(\underline{y}=0)\equiv B_{ij}$. On the other hand, $a_{i}(\underline{y}=0)=0$ is possible, and we  include the first order:
\begin{equation}
a_i(\underline{y})\approx a_i(\underline{y}=0)+y^j\partial_ja_i(\underline{y})\big|_{\underline{y}=0}  \,.
\end{equation}
Defining $A_{ij}=\partial_ja_i(\underline{y})\big|_{\underline{y}=0}$ and $a_i\equiv a_i(\underline{y}=0)$,  the Fokker-Planck equation~(\ref{fokker_planck}) turns into
\begin{align}
\partial_tP(\underline{y},t)=&-a^i\partial_iP(\underline{y},t)-\partial_i\big[A_{ij}y^jP(\underline{y},t)\big] \cr
&+\frac{1}{2}B_{ij}\partial_i\partial_jP(\underline{y},t) \, .
\label{expanded_fokker_planck}
\end{align}

The above expanded Fokker-Planck equation may be solved by the ansatz
\begin{equation}
P(\underline{y},t)\sim \exp\big[-{\textstyle \frac{1}{2}}(\Sigma^{-1})_{ij}y^iy^j+\xi_i^{-1}y^i \big]   \, ,
\end{equation}
which is the generic form of the domain wall resp. boundary layer distribution.
When put into Eq.~(\ref{expanded_fokker_planck}), it leads to a matrix equation for $\Sigma$:
\begin{equation}
\Sigma A^T+A\Sigma+B=0  \,,
\label{Sigma_eq}
\end{equation}
the solution is given in App.~\ref{app_fokker_planck}.
$\xi$ is then determined by
\begin{equation}
\xi^{-1}_i=-(\Sigma^{-1})_{ij}A^{-1}_{jk}a_k \,.
\label{xi}
\end{equation}

Here, we continue by discussing the physical meaning of $\Sigma$ and $\xi$. First,  assume that a domain wall emerges in the density profile of the spin-up states. Then, $\bar{x}_w^\uparrow$ lies in bulk, and we have $a_\uparrow=0$. According to  Eq.~(\ref{xi}),
this implies $\xi_\uparrow^{-1}=0$. The domain wall distribution in the spin-up density is therefore Gaussian and its
width given by $(\Sigma_{\uparrow\uparrow})^{1/2}$.

Second, for a boundary layer forming in the density of spin-up states, it follows that $a_\uparrow\neq 0$ and
therefore also $\xi_\uparrow^{-1}\neq 0$. As for large enough systems fluctuations are small ($y^\uparrow,y^\downarrow\ll 1$), the contribution to the domain wall
distribution coming from $\xi_\uparrow^{-1}y^\uparrow$ is the dominating one. The value $\xi_\uparrow$ thus
describes how far the boundary layer extends into bulk, we refer to it as the \emph{localization length}.\\

\subsection{Approaching the coexistence line}

When the delocalization line is approached, we know from the MF analysis that  domain walls delocalize. In the domain wall picture, this must result in a diverging width of the domain wall resp. boundary layer distributions. In this section, we want to show that this divergence indeed  emerges within the above description. Also, we  calculate the corresponding exponents and compare our findings to stochastic simulations.

We define the quantities
\begin{align}
d^\uparrow=&\beta_\text{eff}^\uparrow(1-\beta_\text{eff}^\uparrow)-\alpha_\text{eff}^\uparrow(1-\alpha_\text{eff}^\uparrow)+\Omega\int_0^1[\rho^\uparrow-\rho^\downarrow]dx\,, \cr
d^\downarrow=&\beta_\text{eff}^\downarrow(1-\beta_\text{eff}^\downarrow)-\alpha_\text{eff}^\downarrow(1-\alpha_\text{eff}^\downarrow)+\Omega\int_0^1[\rho^\downarrow-\rho^\uparrow]dx\,,
\label{dist}
\end{align}
as a measure of distance to the delocalization transition line, where  $d^\uparrow=d^\downarrow=0$ is encountered. Our aim is to calculate $\Sigma$ and $\xi$ for  $d^\uparrow, d^\downarrow\rightarrow 0$, i.e. along a path that ends at the coexistence line.
     
Different paths are possible. Applying the two symmetries of the model, they fall into two classes. First, we may have a localized domain wall in the density of one of the internal states, and a pure LD or HD phase for the other one. In the second class, the densities of both states are in pure LD or HD phases.

Note that the second case is similar to the situation in TASEP. There as well we may approach the coexistence line on a path along which LD or HD is encountered; in this situation, the localization length $\xi$  diverges. In our model, exactly the same behavior emerges: the localization lengths $\xi^\uparrow,\xi^\downarrow$, describing the boundary layers in the densities of spin-up resp. spin-down state, both diverge.

The first case, corresponding to the path shown in red in Fig.~\ref{phasediag_ex}, does not possess an analogy to earlier studied ASEP models. Also off  the delocalization transition line, we have a domain wall, which is localized. It allows us to study how its width diverges when approaching the IN-EX-boundary, i.e. how its delocalization arises.

We start our considerations with the first case. As a representative example, a situation with a localized domain wall in the spin-up  density and a pure LD or HD phase for spin-down is studied. This case implies $d^\uparrow=0$ and $d^\downarrow=\beta_\text{eff}^\uparrow(1-\beta_\text{eff}^\uparrow)+\beta_\text{eff}^\downarrow(1-\beta_\text{eff}^\downarrow)-\alpha_\text{eff}^\uparrow(1-\alpha_\text{eff}^\uparrow)-\alpha_\text{eff}^\downarrow(1-\alpha_\text{eff}^\downarrow)\neq0$. Both
$d^\downarrow<0$ and $d^\downarrow>0$ are possible.
Our focus is on the width $\sigma=(\Sigma_{\uparrow\uparrow})^{1/2}$ of the domain wall distribution in the spin-up density. More specific, we aim at finding the scaling behavior of $\sigma$ depending on the system size $L$ and the distance $d^\downarrow$. We expect that $\sigma\rightarrow 0$ for increasing system size $L\rightarrow\infty$ on the one hand, and that $\sigma$ diverges for $d^\downarrow\rightarrow 0$ on the other hand.

In App.~\ref{app_fokker_planck}, we solve the expanded Fokker-Planck equation~(\ref{expanded_fokker_planck}).  Considering the solution~(\ref{sigma}), we
recognize that $\Sigma\sim(\det{A})^{-1}$, and from~(\ref{detA}) we infer $\det{A}\sim d^\downarrow$.
Combining these two results, we arrive at $\sigma=(\Sigma_{\uparrow\uparrow})^{1/2}\sim |d^\downarrow|^{-1/2}$, such that $\sigma$ diverges when the delocalization transition line is approached. The mathematical reason is that one eigenvalue of $A$ goes to zero (such that $\det{A}\rightarrow 0$). Of course, the eigenvector to the vanishing eigenvalue corresponds to the direction along the one-parameter curve on which the mean-field values $\bar{x}_w^\uparrow,\bar{x}_w^\downarrow$ can reside when the delocalization transition line is reached. Thus, the divergence of the width of the domain wall
distribution originates in the one degree of freedom which exists at the delocalization transition.

The dependence of $\sigma$ on the system size $L$ is also of interest. Remembering $A\sim L^{-1}$ and $B\sim L^{-2}$, and with help of Eq.~(\ref{sigma}),  we arrive  at $\sigma\sim L^{-1/2}$. Thus, with increasing $L$, the domain wall distribution gets sharper aligned to its mean value.
Together with the previous result, we obtain the following scaling behavior for the width $\sigma=(\Sigma_{\uparrow\uparrow})^{1/2}$ of the domain wall distribution in the spin-up density:
\begin{equation}
\sigma\sim|Ld^\downarrow|^{-1/2}  \,.
\end{equation}
Comparing these results to stochastic simulations, see Fig.~\ref{scaling}, we recover a good
agreement.

In the second case, the densities of both states are in pure LD or HD phases, i.e. boundary layers occur in both density profiles. As discussed at the end of Subsec.~\ref{subsec_fokker_planck},  the latter implies $a_\uparrow, a_\downarrow\neq0$, thus $\xi_\uparrow, \xi_\downarrow\neq0$ and $\xi_\uparrow, \xi_\downarrow$ are found to be the localization lengths, describing  how far the boundary layers extend into bulk. We are thus interested in their scaling behavior.

Both $d^\uparrow,d^\downarrow$ are different from $0$; approaching the delocalization transition line translates into $d^\uparrow,d^\downarrow\rightarrow 0$.
Our starting point is again the solution to the expanded Fokker-Planck equation~(\ref{expanded_fokker_planck})  given in App.~\ref{app_fokker_planck}.  From Eq.~(\ref{sigma}) we infer that $\Sigma\sim(\det{A})^{-1} $. Together with $a_i\sim d^i$ and Eq.~(\ref{xi}),  this implies for the localization lengths $\xi_i\sim (d^i)^{-1}$. For the scaling behavior in $L$, we observe $a_i\sim L^{-1}$ and find $\xi_i\sim L^{-1}$. Together, the scaling behavior is given by
\begin{align}  
\xi_\uparrow\sim&(Ld^\uparrow)^{-1}\,,\cr
\xi_\downarrow\sim&(Ld^\downarrow)^{-1}   \, .
\end{align}

As discussed in the beginning of this subsection,  the same scaling behavior  emerges in  TASEP for the localization length $\xi$ of the boundary layer when approaching the coexistence line.

\section{Summary}

The appearance of a domain wall delocalization is a particular feature of  the exclusion process with internal states introduced in Refs.~\cite{reichenbach-2006-97,reichenbach-2007-9}.  Indeed, in the simplest driven exclusion process, the TASEP~\cite{krug-1991-76,derrida-1998-301}, such a delocalization of a localized domain wall does not emerge. There, boundary layers characterize the different phases. A delocalized domain wall forms at the phase boundary between low- and high-density phases, while localized domain walls do not build up.   Upon coupling to external reservoirs~\cite{parmeggiani-2003-90,parmeggiani-2004-70}, localized domain walls may appear in bulk, separating a low-density from a high-density region. Varying the system's parameters, their location changes continuously, and they may leave bulk through the left or the right boundary, causing continuous transitions to pure LD or HD phases. Domain wall delocalization, corresponding to a discontinuous transition, does not emerge.

The exclusion processes with internal states exhibits, in addition to continuous phase transitions similar to the ones described above, \emph{discontinuous transitions}. In particular, in phase space, two regions with localized domain walls may be separated by such a discontinuous transition; there, the domain wall position jumps from a certain location in bulk to another. This jump is connected to a delocalization of the domain wall: Approaching the discontinuous transition, the domain wall delocalizes and, upon crossing the transition, re-localizes at the other position.

We have presented a detailed study of this delocalization phenomenon. First, we have investigated the system's behavior at the discontinuous transition. From mean-field considerations, delocalized domain walls have been identified which perform coupled random walks, resulting in smoothened density profiles that lack sharp shocks. Second, we have investigated the delocalization of a localized domain wall upon approaching the discontinuous transition.  Using the domain wall picture, we have set up a quantitative description of the delocalization. Starting from coupled random walks of the domain walls, we have expanded the corresponding Master-equation in the Kra-mers-Moyal formalism, used the linear noise approximation by van Kampen and obtained an analytically solvable Fokker-Planck equation. The latter has revealed power-law dependences of the domain wall width $\sigma$ on the distance $d$ to the delocalization transition and the system size $L$: $\sigma\sim|Ld|^{-1/2}$.  These findings have been validated by stochastic simulations.

As in similar driven diffusive systems where a localized domain wall appears~\cite{parmeggiani-2003-90,parmeggiani-2004-70}, the domain wall's width tends to zero with increasing system size, and is proportional to its inverse square root. However, in the present system, the domain wall can delocalize, namely upon approaching the discontinuous transition. There, the width diverges, being proportional to the inverse square root of the distance to the transition. 
 
We believe that the above discussed domain wall delocalization  represents a robust and generic phenomenon that may  emerge in other driven diffusive system a well, such as weakly coupled antiparellel transport~\cite{juhasz-2007-76} or in models for intracellular transport on multiple lanes that take motors' internal states into account~\cite{basu-2007-75,chowdhury-2007}. Its identification in other non-equilibrium systems will shed further light on their universal phenomenology.  Its observation in real systems, such as intracellular transport of kinesins moving on the parallel protofilaments of a microtubulus, constitutes a challenge for future research.

Financial support of the German Excellence Initiative via the program ``Nanosystems
Initiative Munich" and the German Research Foundation via the SFB TR12 ``Symmetries and Universalities in Mesoscopic Systems''  is gratefully acknowledged.
Tobias Reichenbach acknowledges funding by the Elite-Netzwerk Bayern.

\label{concl}

\begin{appendix}

\section{The Fokker-Planck equation and its solution \label{app_fokker_planck} }

In this Appendix, we want to give more details concerning the technical parts of Sec.~\ref{domain_wall_pict}. The coefficients of the Fokker-Planck equation~(\ref{fokker_planck}) are derived, and the solution to the resulting matrix equation~(\ref{Sigma_eq}) is given.

\subsection{The coefficients of the Fokker-Planck equation}

The Fokker-Planck equation for the distribution of the domain wall positions reads
\begin{equation}
\partial_tP(\underline{y},t)=-\partial_i[a_i(\underline{y})P(\underline{y},t)]+\frac{1}{2}B_{ij}\partial_i\partial_jP(\underline{y},t) \,.
\end{equation}
The coefficients $a_i$ and $B_{ij}$ are given by  Eqs.~(\ref{a_B}), we exemplify their calculation for $a_\uparrow$. According to~(\ref{a_B}), the latter is connected to the changes $y^\uparrow$,  arising from the processes (i), (ii), (v) and (vi) (see Subsec.~\ref{subsec_fokker_planck}). The changes together with the respective rates are read off from Tab.~\ref{random_walk}. We obtain
\begin{align}
a_\uparrow(\underline{y})=&L^{-1}\Delta_\uparrow^{-1}\Big[-\alpha_\text{eff}^\uparrow(1-\alpha_\text{eff}^\uparrow)
+\beta_\text{eff}^\uparrow(1-\beta_\text{eff}^\uparrow)\cr
&+\Omega\int_0^1\rho^\uparrow[1-\rho^\downarrow]dx
-\Omega\int_0^1\rho^\downarrow[1-\rho^\uparrow]dx\Big]\cr
=&L^{-1}\Delta_\uparrow^{-1}d^\uparrow  \, ,
\end{align}
where we used the definitions~(\ref{dist}) for the distances $d^{\uparrow,\downarrow}$ to the delocalization transition. The other coefficients are obtained along the same lines; together, they read
\begin{align}
a_\uparrow(\underline{y})=&L^{-1}\Delta_\uparrow^{-1}d^\uparrow \,,\cr
a_\downarrow(\underline{y})=&L^{-1}\Delta_\downarrow^{-1}d^\downarrow\,, \cr
B_{\uparrow\uparrow}(\underline{y})=&L^{-2}\Delta_\uparrow^{-2}\Big\{\alpha^\uparrow(1-\alpha^\uparrow)+\beta^\uparrow(1-\beta^\uparrow)+\Omega j^{\text{tot}}\cr
&\quad+\Omega\int_0^1[\rho^\uparrow(x)-\rho^\downarrow(x)]^2dx\Big\} \,, \cr
B_{\uparrow\downarrow}(\underline{y})=&-L^{-2}\Delta_\uparrow^{-1}\Delta_\downarrow^{-1}\Big\{\Omega j^{\text{tot}}\cr &\quad+\Omega\int_0^1[\rho^\uparrow(x)-\rho^\downarrow(x)]^2dx\Big\}\,,  \cr
B_{\downarrow\downarrow}(\underline{y})=&L^{-2}\Delta_\downarrow^{-2}\Big\{\alpha^\downarrow(1-\alpha^\downarrow)+\beta^\downarrow(1-\beta^\downarrow)+\Omega j^{\text{tot}}\,,\cr
&\quad+\Omega\int_0^1[\rho^\uparrow(x)-\rho^\downarrow(x)]^2dx\Big\}   \, .
\end{align}
For the expanded Fokker-Planck equation~(\ref{expanded_fokker_planck}),  the matrix elements $A_{ij}$ are given as the derivatives $A_{ij}=\partial_ja_i(\underline{y})\big|_{\underline{y}=0}$:
\begin{align}
A_{\uparrow\uparrow}=&-L^{-1}\Delta_\uparrow^{-2}(\partial_\uparrow\Delta_\uparrow)d^\uparrow-L^{-1}\,,\cr
A_{\uparrow\downarrow}=&-L^{-1}\Delta_\uparrow^{-2}(\partial_\downarrow\Delta_\uparrow)d^\uparrow+L^{-1}\Delta_\uparrow^{-1}\Delta_\downarrow\,,\cr
A_{\downarrow\uparrow}=&-L^{-1}\Delta_\downarrow^{-2}(\partial_\uparrow\Delta_\downarrow)d^\downarrow+L^{-1}\Delta_\downarrow^{-1}\Delta_\uparrow\,,\cr
A_{\downarrow\downarrow}=&-L^{-1}\Delta_\downarrow^{-2}(\partial_\downarrow\Delta_\downarrow)d^\downarrow-L^{-1}   \, .
\end{align}

The key point of the analysis is the observation that in the determinant of $A$,
\begin{align}
\det&{A}=L^{-2}\big\{
\Delta_\uparrow^{-1}\big[\Delta_\uparrow^{-1}(\partial_\uparrow\Delta_\uparrow)+\Delta_\downarrow^{-1}(\partial_\downarrow\Delta_\uparrow)\big]d^\uparrow\cr
&+\Delta_\downarrow^{-1}\big[\Delta_\downarrow^{-1}(\partial_\downarrow\Delta_\downarrow)+\Delta_\uparrow^{-1}(\partial_\uparrow\Delta_\downarrow)\big]d^\downarrow  \cr
&+\Delta_\uparrow^{-2}\Delta_\downarrow^{-2}\big[(\partial_\uparrow\Delta_\uparrow)(\partial_\downarrow\Delta_\downarrow)-(\partial_\downarrow\Delta_\uparrow)(\partial_\uparrow\Delta_\downarrow)\big]d^\uparrow d^\downarrow
\big\}\,,\cr
\label{detA}
\end{align}
the terms independent of $d^\uparrow,d^\downarrow$ have canceled. Thus, for $d^\uparrow,d^\downarrow\rightarrow 0$, we encounter $\det{A}\rightarrow 0$. We show in the following that this causes the domain walls to delocalize when  $d^\uparrow,d^\downarrow\rightarrow 0$.

\subsection{Solution of the Fokker-Planck equation}

For the  solution of the expanded Fokker-Planck equation~(\ref{expanded_fokker_planck}), we already anticipated the form
\begin{equation}
P(\underline{y},t)\sim \exp\big[-{\textstyle \frac{1}{2}}(\Sigma^{-1})_{ij}y^iy^j+\xi_i^{-1}y^i \big]   \, .
\end{equation}
To solve the resulting matrix equation
\begin{equation}
\Sigma A^T+A\Sigma+B=0\,,
\end{equation}
for $\Sigma$, we write it in vector form:
\begin{equation}
\left(\begin{array}{ccc}
2A_{\uparrow\uparrow} & 2A_{\uparrow\downarrow} &0\\
A_{\downarrow\uparrow} & A_{\uparrow\uparrow}+A_{\downarrow\downarrow} & A_{\uparrow\downarrow} \\
0 & 2A_{\downarrow\uparrow} & 2A_{\downarrow\downarrow}
\end{array} \right)
\left(\begin{array}{c}
\Sigma_{\uparrow\uparrow} \\
\Sigma_{\uparrow\downarrow}\\
\Sigma_{\downarrow\downarrow}
\end{array} \right)
+
\left(\begin{array}{c}
B_{\uparrow\uparrow} \\
B_{\uparrow\downarrow}\\
B_{\downarrow\downarrow}
\end{array} \right)
=0\,.
\end{equation}
Note that $B$ and $\Sigma$ are symmetric.
The above equation can be inverted to give
\begin{align}
&\left(\begin{array}{c}
\Sigma_{\uparrow\uparrow} \\
\Sigma_{\uparrow\downarrow}\\
\Sigma_{\downarrow\downarrow}
\end{array} \right)
=-\frac{1}{2(A_{\uparrow\uparrow}+A_{\downarrow\downarrow})\det{A}}\times \cr
&\times\left(\begin{array}{ccc}
\det{A}+A_{\downarrow\downarrow}^2 & -2A_{\uparrow\downarrow}A_{\downarrow\downarrow} &A_{\uparrow\downarrow}^2\\
-A_{\downarrow\uparrow}A_{\downarrow\downarrow} & 2A_{\uparrow\uparrow}A_{\downarrow\downarrow} & -A_{\uparrow\uparrow}A_{\uparrow\downarrow} \\
A_{\downarrow\uparrow}^2 & -2A_{\uparrow\uparrow}A_{\downarrow\uparrow} & \det{A}+A_{\uparrow\uparrow}^2
\end{array} \right)
\left(\begin{array}{c}
B_{\uparrow\uparrow} \\
B_{\uparrow\downarrow}\\
B_{\downarrow\downarrow}
\end{array} \right)\,.\cr
\label{sigma}
\end{align}

For the scaling behavior, we observe $\Sigma\sim \det A$. When the delocalization transition is approached, as $\det A\rightarrow 0$, also $\Sigma\rightarrow 0$. Thus $\Sigma^{-1}$, describing the widths of the domain walls, diverges; and the domain walls delocalize.

\end{appendix}


\begin{thebibliography}{10}

\bibitem{SchmittmannZia}
B.~Schmittmann and R.~K.~P. Zia.
\newblock In C.~Domb and J.~Lebowitz, editors, {\em Phase Transitions and
  Critical Phenomena}, volume~17. Academic Press, London, 1995.

\bibitem{krug-1991-76}
J.~Krug.
\newblock Boundary-induced phase transitions in driven diffusive systems.
\newblock {\em Phys. Rev. Lett.}, 67:1882, 1991.

\bibitem{georgiev-2005-94}
I.~T. Georgiev, B.~Schmittmann, and R.~K.~P. Zia.
\newblock Anomalous nucleation far from equilibrium.
\newblock {\em Phys. Rev. Lett.}, 94:115701, 2005.

\bibitem{macdonald-1968-6}
C.~T. MacDonald, J.~H. Gibbs, and A.~C. Pipkin.
\newblock Kinetics of biopolymerization on nucleic acid templates.
\newblock {\em Biopolymers}, 6:1, 1968.

\bibitem{hirokawa-1998-279}
N.~Hirokawa.
\newblock Kinesin and dynein superfamily proteins and the mechanism of
  organelle transport.
\newblock {\em Science}, 279:519--526, 1998.

\bibitem{Howard}
J.~Howard.
\newblock {\em Mechanics of Motor Proteins and the Cytoskeleton}.
\newblock Sinauer Press, Sunderland, Massachusetts, 2001.

\bibitem{lipowsky-2001-87}
R.~Lipowsky, S.~Klumpp, and T.~M. Nieuwenhuizen.
\newblock Random walks of cytoskeletal motors in open and closed compartments.
\newblock {\em Phys. Rev. Lett.}, 87:108101, 2001.

\bibitem{kruse-2002-66}
K.~Kruse and K.~Sekimoto.
\newblock Growth of fingerlike protrusions driven by molecular motors.
\newblock {\em Phys. Rev. E}, 66:031904, 2002.

\bibitem{klumpp-2003-113}
S.~Klumpp and R.~Lipowsky.
\newblock Traffic of molecular motors through tube-like compartments.
\newblock {\em J. Stat. Phys.}, 113:233, 2003.

\bibitem{klein-2005-94}
G.~A. Klein, K.~Kruse, G.~Cuniberti, and F.~J\"ulicher.
\newblock Filament depolymerization by motor molecules.
\newblock {\em Phys. Rev. Lett.}, 94:108102, 2005.

\bibitem{Hinsch}
H.~Hinsch, R.~Kouyos, and E.~Frey.
\newblock In A.~Schadschneider, T.~P\"oschel, R.~K\"uhne, M.~Schreckenberg, and
  D.~E. Wolf, editors, {\em Traffic and Granular Flow '05}. Springer, 2006.

\bibitem{zutic-2004-76}
I.~$\check{Z}$uti$\acute{\text{c}}$, J.~Fabian, and S.~Das Sarma.
\newblock Spintronics: Fundamentals and applications.
\newblock {\em Rev. Mod. Phys.}, 76:323, 2004.

\bibitem{reichenbach-2006-97}
T.~Reichenbach, T.~Franosch, and E.~Frey.
\newblock Exclusion processes with internal states.
\newblock {\em Phys. Rev. Lett.}, 97:050603, 2006.

\bibitem{helbing-2001-73}
D.~Helbing.
\newblock Traffic and related self-driven many-particle systems.
\newblock {\em Rev. Mod. Phys.}, 73:1067, 2001.

\bibitem{chowdhury-2000-329}
D.~Chowdhury, L.~Santen, and A.~Schadschneider.
\newblock Statistical physics of vehicular traffic and some related systems.
\newblock {\em Phys. Rep.}, 329:199, 2000.

\bibitem{derrida-1998-301}
B.~Derrida.
\newblock An exactly soluble non-equilibrium system: The asymmetric simple
  exclusion process.
\newblock {\em Phys. Rep.}, 301:65--83, 1998.

\bibitem{Schutz}
G.~Sch\"utz.
\newblock In C.~Domb and J.~Lebowitz, editors, {\em Phase Transitions and
  Critical Phenomena}, volume~19, pages 3--251. Academic Press, San Diego,
  2001.

\bibitem{derrida-1992-69}
B.~Derrida, E.~Domany, and D.~Mukamel.
\newblock An exact solution of a one-dimensional asymmetric exclusion model
  with open boundaries.
\newblock {\em J. Stat. Phys}, 69:667, 1992.

\bibitem{schuetz-1993-72}
G.~M Sch\"{u}tz and E.~Domany.
\newblock Phase transitions in an exactly soluble one-dimensional exclusion
  process.
\newblock {\em J. Stat. Phys.}, 72:277--296, 1993.

\bibitem{derrida-1993-26}
B.~Derrida, M.~R. Evans, V.~Hakim, and V.~Paquier.
\newblock Exact solution of a {1D} asymmetric exclusion model using a matrix
  formulation.
\newblock {\em J. Phys. A: Math. Gen.}, 26:1493, 1993.

\bibitem{kolomeisky-1998-31}
A.~B. Kolomeisky, G.~M. Sch\"{u}tz, E.~B. Kolomeisky, and J.~P. Straley.
\newblock Phase diagram of one-dimensional driven lattice gases with open
  boundaries.
\newblock {\em J. Phys. A: Math. Gen.}, 31(33):6911--6919, 1998.

\bibitem{santen-2002-106}
L.~Santen and C.~Appert.
\newblock The asymmetric exclusion process revisited: Fluctuations and dynamics
  in the domain wall picture.
\newblock {\em J. Stat. Phys}, 106:187, 2002.

\bibitem{reichenbach-2007-9}
T.~Reichenbach, E.~Frey, and T.~Franosch.
\newblock Traffic jams induced by rare switching events in two-lane transport.
\newblock {\em New J. Phys.}, 9:159, 2007.

\bibitem{Mukamel}
D.~Mukamel.
\newblock In M.~Cates and M.~Evans, editors, {\em Soft and Fragile Matter},
  pages 237--258. Institute of Physics Publishing, Bristol, 2000.

\bibitem{parmeggiani-2003-90}
A.~Parmeggiani, T.~Franosch, and E.~Frey.
\newblock Phase coexistence in driven one dimensional transport.
\newblock {\em Phys. Rev. Lett.}, 90:086601, 2003.

\bibitem{parmeggiani-2004-70}
A.~Parmeggiani, T.~Franosch, and E.~Frey.
\newblock The totally asymmetric simple exclusion process with {L}angmuir
  kinetics.
\newblock {\em Phys. Rev. E}, 70:046101, 2004.

\bibitem{schmittmann-2005-70}
B.~Schmittmann, J.~Krometics, and R.~K.~P. Zia.
\newblock Will jams get worse when slow cars move over?
\newblock {\em Europhys. Lett.}, 70:299--305, 2005.

\bibitem{pronina-2004-37}
E.~Pronina and A.~B. Kolomeisky.
\newblock Two-channel totally asymmetric simple exclusion processes.
\newblock {\em J. Phys. A: Math. Gen.}, 37:9907, 2004.

\bibitem{pronina-2006-372}
E.~Pronina and A.~B. Kolomeisky.
\newblock Asymmetric coupling in two-channel simple exclusion processes.
\newblock {\em Phys. A}, 372:12--21, 2006.

\bibitem{gillespie-1976-22}
D.~T. Gillespie.
\newblock Stochastic simulations of chemical processes.
\newblock {\em J. Comp. Phys.}, 22:403, 1976.

\bibitem{gillespie-1977-81}
D.~T. Gillespie.
\newblock Exact simulations of coupled chemical reactions.
\newblock {\em J. Phys. Chem.}, 81:2340--2361, 1977.

\bibitem{pierobon-2005-72}
P.~Pierobon, A.~Parmeggiani, F.~{von Oppen}, and E.~Frey.
\newblock Dynamic correlation functions and {B}oltzmann {L}angevin approach for
  driven one dimensional lattice gas.
\newblock {\em Phys. Rev. E}, 72:036123, 2005.

\bibitem{taeuber-notes}
U.~T\"auber.
\newblock lecture note.

\bibitem{VanKampen}
N.G. van Kampen.
\newblock {\em Stochastic Processes in Physics and Chemistry}.
\newblock North Holland Publishing Company, first edition, 1981.

\bibitem{juhasz-2007-76}
R.~Juhasz.
\newblock Weakly coupled, antiparallel, totally asymmetric simple exclusion
  processes.
\newblock {\em Phys. Rev. E}, 76:021117, 2007.

\bibitem{basu-2007-75}
A.~Basu and D.~Chowdhury.
\newblock Traffic of interacting ribosomes on {mRNA} during protein synthesis:
  effects of chemo-mechanics of individual ribosomes.
\newblock {\em Phys. Rev. E}, 75:021902, 2007.

\bibitem{chowdhury-2007}
D.~Chowdhury and J.~Wang.
\newblock Traffic of single-headed motor proteins {KIF1A}: effects of lane
  changing, 2007.
\newblock arXiv.org:0712.4304.

\end{thebibliography}

\end{document}